\documentclass[a4paper,10pt,onecolumn,fleqn,oneside]{article}%

\usepackage{graphicx}

\usepackage[sort&compress]{natbib}
\date{}
\usepackage{amsmath}%
\usepackage{subfig}%
\newcommand{\mb}[1]{\mathbf{#1}}%
\begin{document}

\title{Validity of Viscous Core Correction Models for Self-Induced Velocity Calculations\footnote{Submitted for review to the Journal of the American Helicopter Society on April 17, 2010. Accepted for publication in the Journal of the American Helicopter Society on July 28, 2011.}}

\author{Wim R. M. Van Hoydonck\\ \small{National Aerospace Laboratory (NLR)}\\\small{1059 CM Amsterdam}\\\small{The Netherlands}\\\small{email: {\tt wim.van.hoydonck@nlr.nl}}\and Michel J. L. van Tooren\\ \small{Faculty of Aerospace Engineering}\\\small{Delft University of Technology}\\\small{2629 HS Delft}\\\small{The Netherlands}}

\maketitle

\begin{abstract}%
    Viscous core correction models are used in free wake simulations to remove the infinite velocities at the vortex centreline. It will be shown that the assumption that these corrections converge to the Biot-Savart law in the far field is not correct for points near the tangent line of a vortex segment. Furthermore, the self-induced velocity of a vortex ring with a viscous core is shown to converge to the wrong value. The source of these errors in the model is identified and an improved model is presented that rectifies the errors. It results in correct values for the self-induced velocity of a viscous vortex ring and induced velocities that converge to the values predicted by the Biot-Savart law for all points in the far field.
\end{abstract}
\section{Introduction}%
    In engineering rotorcraft simulations such as Refs.~\cite{sadler_1972_1} and~\cite{bagai_leishman_1995}, the wake generated by the rotor blades is often represented by a series of connected straight vortex line segments. The total velocity induced by the wake at points in the flow field is then found by summing the contributions of each individual segment, for which an analytical solution of the Biot-Savart law is readily available (such as Ref.~\cite{katz_plotkin_book_2001}). More sophisticated approaches, such as Johnson (Ref.~\cite{johnson_camrad_1980}) represent part of the wake with quadrilateral panels or, as in Ref.~\cite{bliss_teske_quackenbush_1987} with parabolic arcs. For a point close to a particular line segment, the induced velocity value calculated with the Biot-Savart law is unrealistically high. In literature, these large velocities are corrected by regularising the Biot-Savart law with a model that represents the viscous effects inside the vortex core. One of the first researchers to introduce such a model was Scully (Ref.~\cite{scully_1975}). For points located on the centerline of a line segment, the induced velocity contribution of that segment is undefined in the potential case and zero for the regularised model. For a curved filament discretised using straight segments, this means that the self-induced velocity due to local curvature is not taken into account. Several methods are used in literature to account for this small but nonzero contribution. Scully (Ref.~\cite{scully_1975}) uses analytical expressions for the self-induced velocity of a curved filament, while in other codes a cut-off method is used (Ref.~\cite{bliss_teske_quackenbush_1987}). A third option is to take no special measures for the self-induced velocity of a curved filament (Ref.~\cite{bhagwat_leishman_2001_2}).
    
    In this note, the validity of viscous corrections as found in literature will be analyzed for the case of a vortex ring discretized with a set of straight line segments and for the case where it is represented exactly using a parametric curve.
\section{Filament Induced Velocity Calculations}\label{sec:ind_vel_calcs_using_biot_savart_law}%
\begin{figure}
    \centering
    \subfloat[][]{\label{subfig:straight_segm_core_correction_bug}\includegraphics{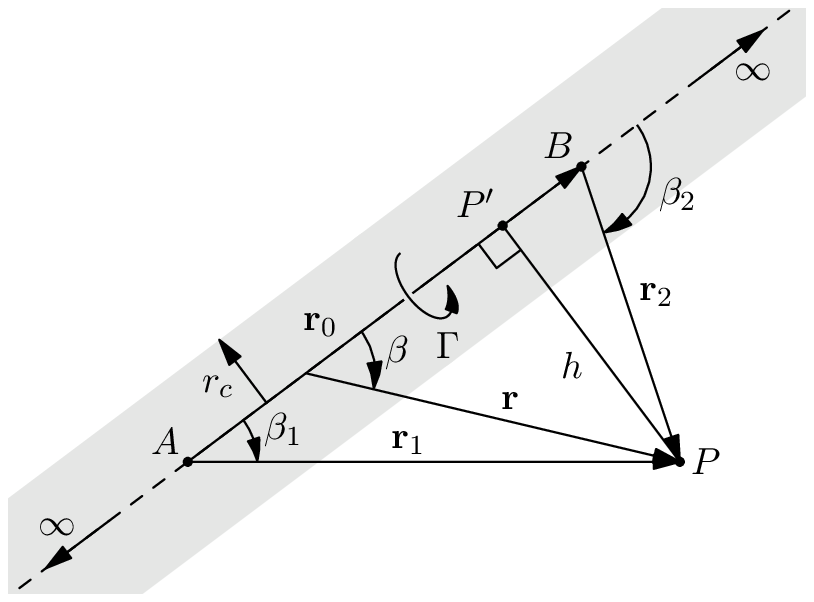}}\hspace{1cm}%
    \subfloat[][]{\label{subfig:curved_segm_core_correction_bug}\includegraphics{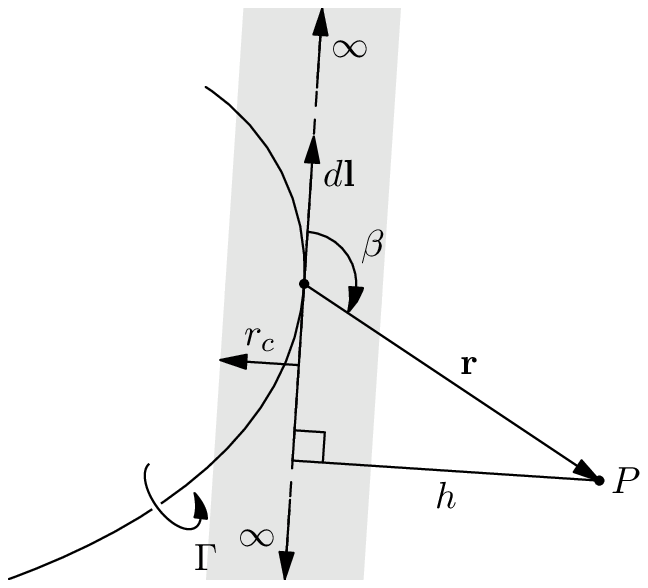}}%
    \caption{Definition of view angles as used in the evaluation of the induced velocity of a vortex filament using the Biot-Savart law. The regularisation region (grey area) of the core model corrections for a straight~\protect\subref{subfig:straight_segm_core_correction_bug} and a curved~\protect\subref{subfig:curved_segm_core_correction_bug} vortex filament as used in literature is also shown.}\label{fig:biot_savart_view_angles_core_region}
\end{figure}

    For a straight vortex line between point $A$ and $B$, the induced velocity at a point $\mb{r}_P$ can be calculated with (Ref.~\cite{katz_plotkin_book_2001}, p. 39),
    \begin{equation}\label{eq:biot_savart_angles}
        \mb{v} = \frac{\Gamma}{4\pi h} \int_{\beta_1}^{\beta_2} \sin \beta d\beta\:\mb{e} = \frac{\Gamma}{4\pi h} (\cos \beta_1 - \cos \beta_2)\:\mb{e}
    \end{equation}
    where the angles $\beta_1$ and $\beta_2$ are called the view angles, $\mb{e}$ is the direction vector of the induced velocity and $h$ is the perpendicular distance between the vortex line and point $P$ (see Fig.~\ref{subfig:straight_segm_core_correction_bug}).
    
    When a vortex filament can be represented with a parametric curve $C(u)$, the Biot-Savart law can be written as
    \begin{equation}\label{eq:biot_savart_param_eqn}
        \Delta \mb{v}(u) = \frac{\Gamma}{4\pi} \frac{d\mb{l}(u) \times \mb{r}(u)}{|\mb{r}(u)|^3}
    \end{equation}
    where $d\mb{l}(u) = dC(u)/du$, the derivative of the curve at the parametric point $u$. Then, for $m$ quadrature points $\mb{u}_i$ on the filament the induced velocity increment $\Delta \mb{v}_i$ is calculated, converting the Biot-Savart integral into a finite sum,
    \begin{equation}\label{eq:quad_rule_sum}
        \mb{v} = \sum_{i=1}^{m} w_i \Delta \mb{v}_i
    \end{equation}
    with $w_i$ the weights of the quadrature rule.
    
\begin{figure}
    \centering%
    \subfloat[][Evaluation points on the $X$-axis where the induced velocity is calculated.]{%
        \label{subfig:vortex_ring_target_point_positions}%
        \resizebox{.45\textwidth}{!}{\includegraphics{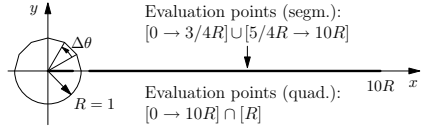}}}\\%
    \subfloat[][$\log_{10}$ of relative error of induced velocity for the parametric vortex ring method for Gauss-Legendre quadrature rules with $m$ abscissae.]{%
        \label{subfig:vortex_ring_rel_error_param}%
        \resizebox{0.45\textwidth}{!}{\includegraphics{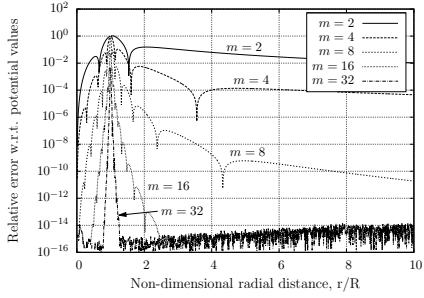}}}\hspace{1cm}%
    \subfloat[][$L_2$-norm of relative error of induced velocity for segmented vortex ring method as a function of the discretisation.]{%
        \label{subfig:vortex_ring_rel_error_segms}%
        \resizebox{0.45\textwidth}{!}{\includegraphics{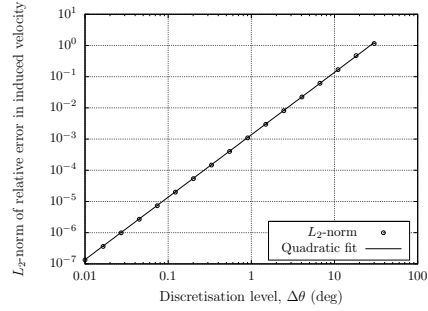}}}%
    \caption{Convergence of induced velocity calculations of quadrature-based method~\protect\subref{subfig:vortex_ring_rel_error_param}  and segmentation method~\protect\subref{subfig:vortex_ring_rel_error_segms} to potential reference values for evaluation points in the plane of a vortex ring~\protect\subref{subfig:vortex_ring_target_point_positions}, on the positive X-axis, from $(0,0)$ to $(10R,0)$. Analytical values (Ref.~\cite{castles_leeuw_1954}) are used as reference.}\label{fig:iv_vortex_ring_rel_error}
\end{figure}

    The efficiency and accuracy of these methods can be verified by calculating the induced velocity in the far field of a vortex ring and comparing the results with analytical solutions. For a set of points in the plane of the vortex ring on the $X$-axis (0 to 10 times the vortex radius $R$, see Fig.~\ref{subfig:vortex_ring_target_point_positions}), the induced velocity is calculated and the relative errors\footnote{The relative error is defined as $|$reference-approximation$|$/$|$reference$|$.} w.r.t. the analytical results as given by Castles and De Leeuw (Ref.~\cite{castles_leeuw_1954}) for both methods are displayed in Figs.~\ref{subfig:vortex_ring_rel_error_param} and~\ref{subfig:vortex_ring_rel_error_segms}. 
    
    Results calculated with the quadrature method (Fig.~\ref{subfig:vortex_ring_rel_error_param}) for $m=32$ (for a total of 128 Biot-Savart function evaluations per evaluation point) are accurate to machine precision everywhere in the field except for a small region close to the vortex ring itself ($3/4R \rightarrow 5/4R$). With this number of Biot-Savart law evaluations ($\Delta \theta \approx 3^{\circ}$) for the same evaluation points, the segment line method (Fig.~\ref{subfig:vortex_ring_rel_error_segms}) only reaches an accuracy of approximately $10^{-2}$. The second-order trend for this method as reported in Ref.~\cite{bhagwat_leishman_2001_2} is clearly visible: a ten-fold increase in the number of segments reduces the relative error by a factor 100.
    
    As the distance between evaluation points and the vortex ring is reduced to zero, the induced velocity increases unboundedly. On the vortex ring itself, the induced velocity is undefined. To remove this singularity, core models are introduced that mimic the viscous behaviour of a real vortex. In general, vortex cores have three nonzero velocity components but it is assumed that only the swirl velocity is non-zero. The other two components (axial and radial velocity) are neglected in most engineering wake models.
    
    On page 601 of Ref.~\cite{leishman_book_2006}, Leishman gives a modification of Eq.~(\ref{eq:biot_savart_angles}) that incorporates the Vatistas (Ref.~\cite{vatistas_kozel_mih_1991}) core model,
    \begin{equation}\label{eq:leishman_biot_savart_mod_vatistas}
        \mb{v} = \frac{\Gamma}{4\pi} \frac{h}{(r_c^{2n} + h^{2n})^{1/n}} (\cos \beta_1 - \cos \beta_2) \vec{e}
    \end{equation}
    where $r_c$ is the core size and $n$ is a small integer. Similar modifications to the Biot-Savart law can be derived for other core models, Johnson (Ref.~\cite{johnson_book_1980}) gives the modification for the model of Scully (Ref.~\cite{scully_1975}) on page 543.
    Equation~(\ref{eq:biot_savart_param_eqn}) can be modified in a similar way to remove the singular behaviour at the core centre,
    \begin{equation}\label{eq:biot_savart_diff_core_mod}
        \Delta \mb{v} = \frac{\Gamma}{4\pi} \frac{h^2}{(r_c^{2n} + h^{2n})^{1/n}} \frac{d\mb{l}(u) \times \mb{r}(u)}{|\mb{r}(u)|^3}
    \end{equation}
    
    In Fig.~\ref{fig:biot_savart_view_angles_core_region}, the shaded regions denote the area where induced velocity values are modified by the two-dimensional core correction model. Near the centerline of an element, the modification region extends to infinity due to its two-dimensional nature. The actual width of the correction region depends on the specific core model used, for the Rankine core model it is exactly the same as the core size $r_c$. For other models, the width will be slightly larger than the core size.
    
\begin{figure}
    \centering
    \includegraphics{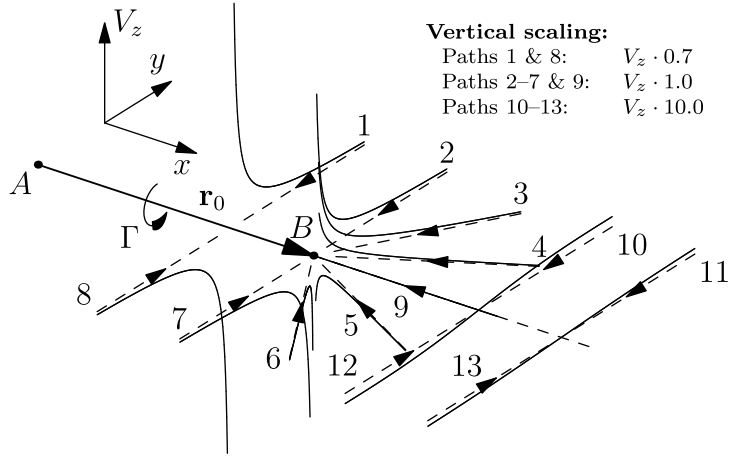}
    \caption{Approaching a vortex segment AB along different paths results in asymptotic behaviour for paths 1 to 8. Due to symmetry, along path 9, the induced velocity is zero everywhere except for the end point and as a results, along paths 10 to 13 no asymptotes are encountered when approaching the centerline.}\label{fig:vortex_line_segment_approach}
\end{figure}

    A vortex segment with end points $A$ and $B$ is shown in Fig.~\ref{fig:vortex_line_segment_approach} together with thirteen paths that approach it. Along path 1 to 8, the induced velocity (calculated with Eq.~(\ref{eq:biot_savart_angles})) increases without bounds. Along path 9 the induced velocity is zero everywhere except for the end point, where it is undefined. As a consequence, at the end of paths 10 to 13 (on the vortex centerline), the induced velocity is zero.
    
\begin{figure}
    \centering
    \subfloat[][]{\label{subfig:straight_segm_core_correction_new}\includegraphics{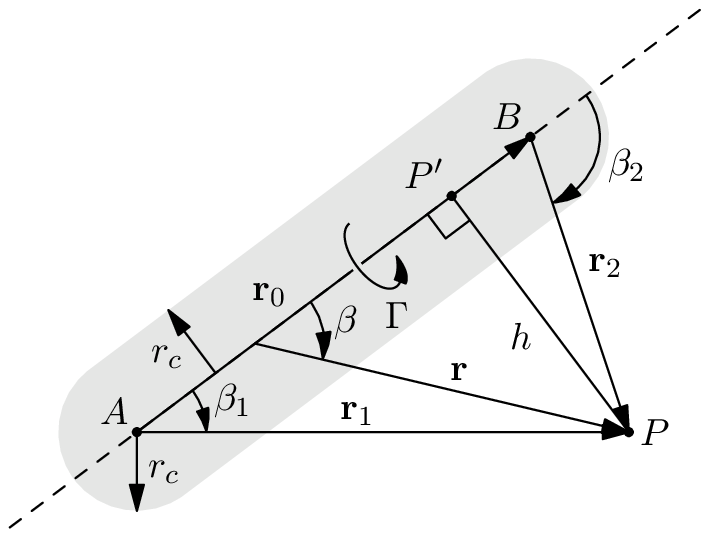}}\hspace{1cm}%
    \subfloat[][]{\label{subfig:curved_segm_core_correction_new}\includegraphics{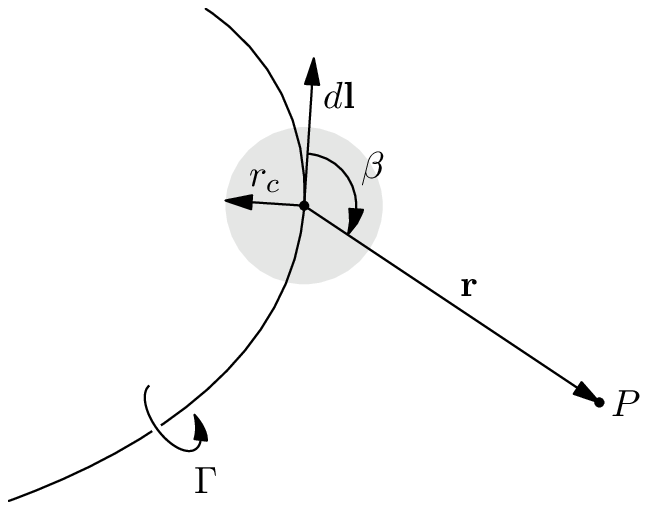}}%
    \caption{Modified core correction model where the modification of the potential induced velocity values is restricted to an area close to the vortex filament for a straight finite length filament~\protect\subref{subfig:straight_segm_core_correction_new} and for a curved filament~\protect\subref{subfig:curved_segm_core_correction_new}.}\label{fig:curved_vortex_iv}
\end{figure}

    Comparison of the correction regions in Fig.~\ref{fig:biot_savart_view_angles_core_region} with the paths along which the induced velocity increases unboundedly (Fig.~\ref{fig:vortex_line_segment_approach}) shows that the unconditional use of the perpendicular distance $h$ in Eqs.~(\ref{eq:leishman_biot_savart_mod_vatistas}) and~(\ref{eq:biot_savart_diff_core_mod}) to correct induced velocity values is wrong. Corrections are applied where no asymptotes are encountered. A simple solution is to check where the projection of point $P$ on line segment AB ($P'$) is located (see Fig.~\ref{subfig:straight_segm_core_correction_new}). If it falls between the end points of the segment, the perpendicular distance is used as before. If the projection lies before point A or after point B, the perpendicular distance $h$ is replaced by the radial distance to the respective end point $|\mb{r}_1|$ or $|\mb{r}_2|$. In terms of the view angles $\beta_1$ and $\beta_2$, Eq.~(\ref{eq:leishman_biot_savart_mod_vatistas}) is modified as follows,
    \begin{equation}\label{eq:leishman_biot_savart_mod_vatistas_new}
        \begin{split}
            \mb{v} &= \frac{\Gamma}{4\pi} \frac{d}{(r_c^{2n} + d^{2n})^{1/n}} (\cos \beta_1 - \cos \beta_2) \vec{e},\\
            &\textrm{where}
            \begin{cases}
                d = |\mb{r}_1| & \textrm{if} \quad \cos \beta_1 < 0,\\
                d = |\mb{r}_2| & \textrm{if} \quad \cos \beta_2 > 0,\\
                d = h & \textrm{elsewhere.}
            \end{cases}
        \end{split}
    \end{equation}

    A comparison of the shaded regions in Figs.~\ref{subfig:straight_segm_core_correction_bug} and~\ref{subfig:straight_segm_core_correction_new} shows that using Eq.~(\ref{eq:leishman_biot_savart_mod_vatistas_new}), the viscous core corrections are restricted to a region close to the vortex segment and do not extend to infinity.
    
\begin{figure}
    \centering
    \subfloat[][Self-induced velocity prediction with original core correction model (Eqs.~(\ref{eq:leishman_biot_savart_mod_vatistas}) and~(\ref{eq:biot_savart_diff_core_mod})).]{\label{subfig:vortex_ring_siv_old_correction}\resizebox{0.45\textwidth}{!}{\includegraphics{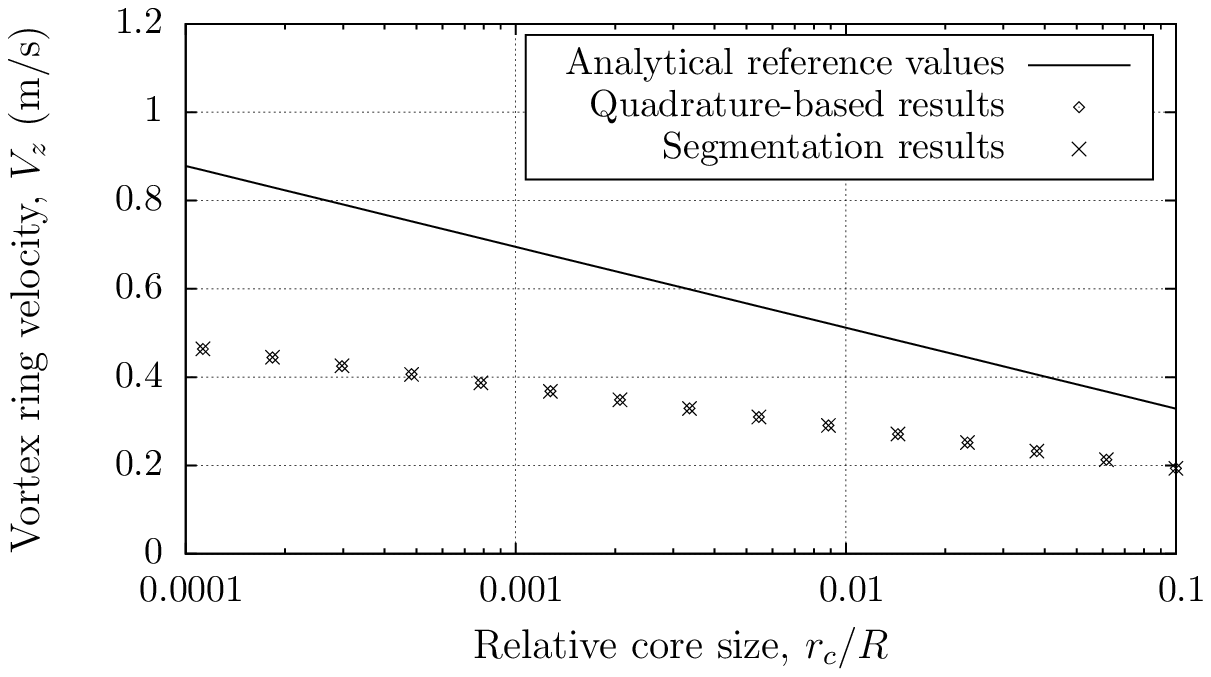}}}\hspace{1cm}
    \subfloat[][Self-induced velocity prediction with modified core correction model (Eqs.~(\ref{eq:leishman_biot_savart_mod_vatistas_new}) and~(\ref{eq:biot_savart_diff_core_mod_new})).]{\label{subfig:vortex_ring_siv_new_correction}\resizebox{0.45\textwidth}{!}{\includegraphics{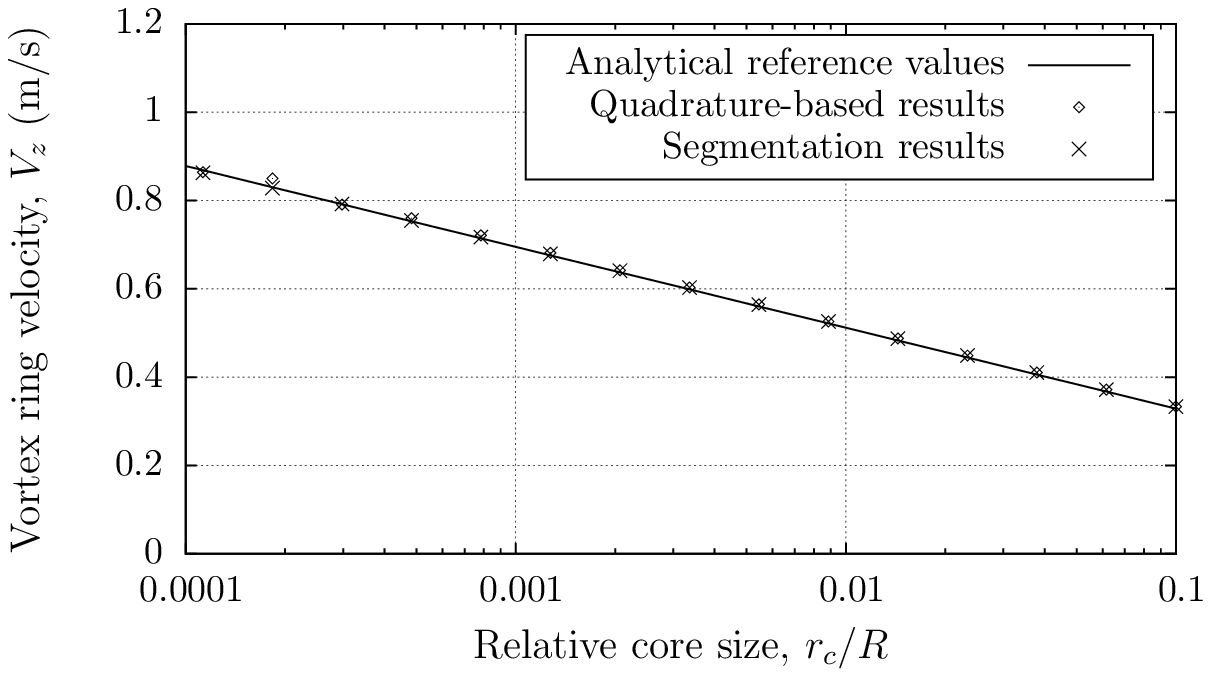}}}
    \caption{Self-induced velocity of a viscous vortex ring as a function of relative core size for the original~\protect\subref{subfig:vortex_ring_siv_old_correction} and modified~\protect\subref{subfig:vortex_ring_siv_new_correction} core correction model.}\label{fig:vortex_ring_siv_core_size_variation}
\end{figure}

    When the length of the vortex segment $|\mb{r}_0|$ is reduced to zero, the distances from the end points to the evaluation point are equal, and the influence region reduces to a circular area (see Fig.~\ref{subfig:curved_segm_core_correction_new}). Equation~(\ref{eq:biot_savart_diff_core_mod}) is modified as follows,
    \begin{equation}\label{eq:biot_savart_diff_core_mod_new}
        \Delta \mb{v} = \frac{\Gamma}{4\pi} \frac{|\mb{r}(u)|^2}{(r_c^{2n} + |\mb{r}(u)|^{2n})^{1/n}} \frac{d\mb{l}(u) \times \mb{r}(u)}{|\mb{r}(u)|^3}.
    \end{equation}
    The influence of these modifications on the prediction of the self-induced velocity of a viscous vortex ring will be assessed here. For a viscous vortex ring with a Rankine core model, Saffman (Ref.~\cite{saffman_book_1992}, p. 195) gives an analytical solution for the self-induced velocity,
    \begin{equation}\label{eq:siv_rankine_vortex_ring}
        V_z = \frac{\Gamma}{4\pi R} \left( \log \frac{8R}{r_c} - \frac{1}{4}\right)
    \end{equation}
    where $r_c/R$ is the ratio of the core size to the vortex ring radius ($r_c/R \ll 1$). For a range of relative core sizes of interest to rotorcraft applications, the self-induced velocity is calculated using the quadrature-based method and the straight-line vortex method as outlined before. The segmentation results were calculated using a circle discretised with 360,000 segments ($\Delta\theta = 0.001^{\circ}$) and the results for the quadrature-based method are generated with Gauss quadrature rules with 128 abscissae.
    Results of these calculations are shown in Fig.~\ref{fig:vortex_ring_siv_core_size_variation}. For the original core correction model (Fig.~\ref{subfig:vortex_ring_siv_old_correction}), it is clear that the self-induced velocity is underpredicted by approximately 44\%, both for the segmentation method and the quadrature method. 
    If Eqs.~(\ref{eq:leishman_biot_savart_mod_vatistas_new}) and~(\ref{eq:biot_savart_diff_core_mod_new}) are used (Fig.~\ref{subfig:vortex_ring_siv_new_correction}), self-induced velocity values are the same as those calculated with Eq.~(\ref{eq:siv_rankine_vortex_ring}).
\section{Conclusions}%
    Using a viscous vortex ring as a basis, it is shown that core correction models used to regularise the Biot-Savart law contain an error caused by the unconditional use of the perpendicular distance in those formulas.
    Self-induced velocities of a viscous vortex ring are shown to converge to the wrong value. The deviation is approximately 44\% from the theoretical value, both for the classical approach where the vortex ring is discretised using straight segments and a numerical quadrature method that uses an exact representation of the vortex ring. A simple modification to the regularised Biot-Savart formula is proposed for both methods that rectifies these errors.

\bibliography{biblio}

\end{document}